\documentclass[12pt,letter]{article}
\pdfoutput=1
\usepackage{graphicx, epsfig, color,cite}
\usepackage{amsmath}
\usepackage{amssymb}
\usepackage{float}
\usepackage{subfig}
\usepackage{hyperref}

\def\to{\rightarrow}

\def\bi{\begin{itemize}}
\def\ei{\end{itemize}}

\def\alt{\lesssim}
\def\agt{\gtrsim}
\def\be{\begin{equation}}  
\def\ee{\end{equation}}  
\def\bea{\begin{eqnarray}}  
\def\eea{\end{eqnarray}}

\begin{document}
\begin{titlepage}
\begin{flushright}
OU-HEP-260703
\end{flushright}

\vspace{0.5cm}
\begin{center}
  {\Large \bf Serendipitous supersymmetric solution to the \\
    strong CP problem
  (and the origin of global $U(1)_{PQ}$)}\\
\vspace{1.2cm} \renewcommand{\thefootnote}{\fnsymbol{footnote}}
{\large Howard Baer$^{1}$\footnote[1]{Email: baer@ou.edu },
Vernon Barger$^2$\footnote[2]{Email: barger@pheno.wisc.edu} and
Dibyashree Sengupta$^{3}$\footnote[3]{Email: sengupta.dibyashree@ucy.ac.cy}
}\\ 
\vspace{1.2cm} \renewcommand{\thefootnote}{\arabic{footnote}}
{\it 
$^1$Homer L. Dodge Department of Physics and Astronomy,
University of Oklahoma, Norman, OK 73019, USA \\[3pt]
}
{\it 
$^2$Department of Physics,
University of Wisconsin, Madison, WI 53706 USA \\[3pt]
}
{\it
$^3$ Department of Physics, University of Cyprus, P.O. Box 20537, 1678 Nicosia, Cyprus} \\[3pt]
\end{center}

\vspace{0.5cm}
\begin{abstract}
\noindent
The Minimal Supersymmetric Standard Model (MSSM) has several problems:
1. its $\mu$ term must be forbidden, then regenerated at the weak scale,
2. it allows for $R$-parity violating superpotential terms which lead to
rapid proton decay, 3. it allows for dimension-5 proton decay operators.
The usual imposition of $R$- or matter parity $P_M$ solves only the second
of these, whereas anomaly-free discrete $\mathbb{Z}_n^R$ symmetries
(consistent with grand unification) address all of them.
Once the $\mu$-term is forbidden by the imposition of a discrete
$\mathbb{Z}_n^R$ symmetry (which can emerge as a discrete remnant of
string compactifications to 4-dimensions),
the MSSM develops an accidental global $U(1)_{PQ}$ symmetry
(thus providing a plausible origin for the global $U(1)_{PQ}$ needed
for solving the strong CP problem).
By coupling the Higgs fields to PQ-charged gauge singlet fields $X,\ Y$
(in the Kim-Nilles mechanism),
and imposing SUSY breaking, one regenerates $\mu$ at the weak scale
whilst breaking the discrete $\mathbb{Z}_n^R$ and the $U(1)_{PQ}$.
The broken global $U(1)_{PQ}$ develops a pseudo-Goldstone boson, the DFSZ axion,
thus (perhaps inadvertently) solving the strong CP problem.
In this setting, SUSY develops a dark matter candidate, the SUSY DFSZ axion,
and possibly, though not necessarily, a WIMP dark matter candidate as well,
depending on the order of the induced $R$-parity violating operators.
\end{abstract}
\end{titlepage}

\section{Introduction}
\label{sec:intro}

The Standard Model (SM) is beset with three finetuning problems
(if we include gravity in the mix):
1. the cosmological constant (CC) problem,
2. the gauge hierarchy problem (GHP, due to Higgs mass instability under radiative corrections) and
3. the strong CP problem (why is the QCD $\bar{\theta}$ parameter
so tiny, $\bar{\theta}\equiv \theta+\arg (\det (M))\lesssim 10^{-10}$, as required
by measurements of the neutron EDM\cite{Abel:2020pzs}?).
The first of these, the CC problem,
does not seem to have a satisfactory solution, although anthropic selection
of tiny $\Lambda$ could occur in the context of a multiverse as expected
from the string landscape\cite{Polchinski:2006gy}.
Many avenues for the GHP have been explored, but the most compelling
is that of weak scale supersymmetry (SUSY), which is also experimentally
well-supported by a variety of precision measurements when compared to
radiative corrections\cite{Baer:2020kwz}.

The most convincing solution to the third of these, the strong CP problem,
seems to be the presence of a spontaneously broken global $U(1)_{PQ}$
symmetry\cite{Peccei:1977hh,Peccei:1977ur,Weinberg:1977ma,Wilczek:1977pj}
which gives rise to a very weakly coupled axion field which allows the
offending CP-violating term in QCD to dynamically relax to (near)
zero\cite{Kim:2008hd,DiLuzio:2020wdo}.
But the origin of the required global $U(1)_{PQ}$ remains a mystery.
From the recent review of Ref. \cite{DiLuzio:2020wdo}, we quote:
\begin{quotation}
  a ‘standard model’ for explaining the origin of the PQ symmetry is still
  missing.$\cdots$
  Candidate models should in first place provide a cogent explanation
  for how a global PQ symmetry arises, and should also naturally embed some
  mechanism to preserve it at an exceptionally good level.
  However, they would acquire real credibility if, with no additional or ad hoc
  theoretical inputs, they could automatically shed light on some other
  unsettled issue of the SM$\cdots$
  We believe that any progress in this direction would represent an
  important milestone to strengthen the plausibility of the axion hypothesis.
\end{quotation}
In this letter, we emphasize that the global $U(1)_{PQ}$ symmmetry and
concommitant solution to the strong CP problem naturally emerge from a
proper implementation of the Minimal Supersymmetric Standard Model which
includes a solution to the SUSY $\mu$ problem\cite{Bae:2019dgg}
via discrete $R$-symmetries
$\mathbb{Z}_n^R$ (which arise from string compactifications from 10 to
4 spacetime dimensions).

\section{SUSY and strong $CP$}

The axion solution to the strong CP problem and the SUSY solution to the GHP
are usually thought to be independent entities,
and lead to different dark matter candidates:
the axion field produced as cold dark matter in the early universe
via coherent field oscillations\cite{Sikivie:2006ni}
(and other mechanisms\cite{Bae:2014rfa}) and the
lightest SUSY particle (LSP) (provided $R$-parity is conserved),
expected to be the lightest neutralino, a
weakly interacting massive particle (WIMP) candidate\cite{Jungman:1995df}.

It is not well-appreciated that a solution to the strong CP problem
actually emerges from the second of these via a proper implementation of the
Minimal Supersymmetric Standard Model (MSSM).
The MSSM begins with the SM gauge symmetry $SU(3)_C\times SU(2)_L\times U(1)_Y$
with fields now elevated to superfields, and where two Higgs doublets
are now required so that higgsino contributions to triangle anomalies will
cancel\cite{Baer:2006rs}.
The superpotential is given by
\bea
W &=& W_{MSSM}+W_\nu+W_{RPV}+W_5\\
W_{MSSM}&=& \mu H_uH_d+f_e^{ij}L_iH_dE_j^c+f_d^{ij}Q_iH_dD+j^c+f_u^{ij}Q_iH_uU_j^c\\
W_\nu &=& f_\nu^{ij}L_iH_uN_j^c +{1\over 2}M_{ij}N_i^cN_j^c\\
W_{RPV} &=& \kappa_i L_iH_h +\lambda_{ijk}L_iL_jE_k^c+\lambda^{\prime}_{ijk}L_iQ_jD_k^c+\lambda^{\prime\prime}_{ijk}U_i^cD_j^cD_k^c\\
W_5 &=& (\kappa_{ijkl}^{(1)}/m_P)Q_iQ_jQ_kL_l +(\kappa_{ijkl}^{(2)}/m_P)U_i^cU_j^cD_k^cE_l^c
\eea
Once the SUSY conserving MSSM Lagrangian is developed (using $W_{MSSM}+W_\nu$),
it is augmented by weak-scale soft SUSY breaking terms.
In gravity-mediation, which is heavily favored by the measured value of the
Higgs mass $m_h\simeq 125$ GeV, SUSY is expected to be broken in a hidden
sector at a scale $m_{hidden}\sim \sqrt{m_{weak}\, m_P}$ so that soft terms
of order $m_{soft}\sim m_{hidden}^2/m_P\sim m_{weak}$ are developed.
The large trilinear $A_t$ soft term helps lift $m_h\to 125$ GeV while
increasing the naturalness of the model\cite{Baer:2024fgd}.
  As a final step, $R$-parity
conservation is invoked on a somewhat ad-hoc basis in order to
eliminate $B$ and $L$ violating terms from the superpotential $W_{RPV}$
which would otherwise lead to rapid proton decay\cite{Dreiner:1997uz}.

\section{Constructing the MSSM: role of $R$-parity}

This {\it usual} procedure for constructing the MSSM actually side-steps
two other important issues:
\bi
\item The SUSY $\mu$ problem which arises because the superpotential
  Higgs bilinear term $W_{MSSM} \supset \mu H_uH_d$ is actually SUSY-conserving.
  The associated mass scale $\mu$ would be expected at $m_P$ rather than
  the phenomenologically required value of $m_{soft}$.
  In order to solve the SUSY $\mu$ problem\cite{Bae:2019dgg}, one must first forbid
  it from the superpotential, and then regenerate it at the weak scale.
  Three ways to regenerate it include 1. the Kim-Nilles (KN)
  mechanism\cite{Kim:1983dt} where $\mu$ arises from non-renormalizable terms in
  the superpotential, 2. the Giudice-Masiero (GM) mechanism\cite{Giudice:1988yz} where $\mu$
  arises from non-renormalizable terms in the K\"ahler potential and 3.
  where $\mu$ arises from the Higgs bilinear coupling to a SM gauge singlet
  superfield as in the NMSSM\cite{Ellwanger:2009dp}.
\item  Dangerous dimension-5 proton decay operators\cite{Hinchliffe:1992ad}
  in $W_5$
  need suppressed coefficients $\kappa_{ijkl}^{(1,2)}\sim 10^{-7}$ in order to
  respect bounds from p-decay limits.
  \ei
  While the invocation of RPC forbids $B$ and $L$ violation in the superpotential,
  it allows for the $\mu$ term and the $W_5$  proton decay terms.

  Ibanez and Ross classified all anomaly-free discrete gauge
  symmetries of order 2 and 3\cite{Ibanez:1991pr};
  they then proposed baryon triality $B_3$ as an alternative.
  Under $B_3$, $B$ violating superpotential terms are forbidden while
  $L$-violating terms are allowed, thus saving the proton from dim-4 decay
  operators. $B_3$ also forbids the $W_5$ p-decay operators, but allows for the $\mu$ term.
  Dreiner, Luhn and Thormeier examined discrete gauge symmetries at higher
  order $N$, and proposed {\it proton hexality} $P_6$ as an alternative to
  RPC\cite{Dreiner:2005rd}.
  Under $P_6$, all $RPV$ superpotential terms are forbidden as are the
  $W_5$ $p$-decay operators.
  The $\mu$ term is still allowed under $P_6$.
  Furthermore, the $P_6$ charge assignments conflict with unified matter
  representations in $SU(5)$ and $SO(10)$ GUTs.
  This is of concern even in the more attractive modern day rendition of
  {\it local} GUTs where matter representations appear at various locales in
  compactified extra-dimensions as found in string constructs\cite{Buchmuller:2005sh,Ratz:2007my,Nilles:2009yd}.

\section{Virtues of discrete $R$-symmetries}
\label{sec:Rsym}

In a remarkable paper, Lee {\it et al.}\cite{Lee:2011dya} catalogued
all orders of discrete $R$-symmetries that are anomaly-free
(up to a universal Green-Schwarz term\cite{Green:1984sg}) and with
$R$-charge values consistent with $SU(5)$ or $SO(10)$ unification.  

$R$-symmetries (for a brief review, see {\it e.g.} \cite{Martin:1997ns})
are characterized by the fact that superspace coordinates 
$\theta$ carry non-trivial $R$-charge: 
in the simplest case, $Q_R(\theta )=+1$ so that $Q_R( d^2\theta ) =-2$. 
For the Lagrangian ${\cal L}\ni \int d^2\theta W$ to be invariant under 
$R$-symmetry, then the superpotential $W$ must carry $Q_R(W)= 2$. 
Discrete $R$ symmetries are appealing since rather than being imposed ad-hoc,
they are expected to  emerge as remnants of 10-$d$ Lorentz symmetry under
compactification of extra dimensions in
superstring theory\cite{Lee:2011dya,Nilles:2013lda,Nilles:2017heg}.
The $\mathbb{Z}_n^R$ symmetry gives rise to a universal gauge 
anomaly $\rho$ mod $\eta$ where the remaining contribution $\rho$ 
is cancelled by the Green-Schwarz axio-dilaton shift and
$\eta =N$ ($N/2$) for $N$  odd (even).  
The anomaly free $R$ charges of various MSSM fields are listed in 
Table \ref{tab:R} for $N$ values consistent with grand unification.
\begin{table}[!htb]
\renewcommand{\arraystretch}{1.2}
\begin{center}
\begin{tabular}{c|ccccc}
multiplet & $\mathbb{Z}_{4}^R$ & $\mathbb{Z}_{6}^R$ & $\mathbb{Z}_{8}^R$ & $\mathbb{Z}_{12}^R$ & $\mathbb{Z}_{24}^R$ \\
\hline
$H_u$ & 0  & 4  & 0 & 4 & 16 \\
$H_d$ & 0  & 0  & 4 & 0 & 12 \\
$Q$   & 1  & 5 & 1 & 5  & 5 \\
$U^c$ & 1  & 5 & 1 & 5  & 5 \\
$E^c$ & 1  & 5 & 1 & 5  & 5 \\
$L$   & 1  & 3 & 5 & 9  & 9 \\
$D^c$ & 1  & 3 & 5 & 9  & 9 \\
$N^c$ & 1  & 1 & 5 & 1  & 1 \\
\hline
\end{tabular}
\caption{Derived MSSM field $R$ charge assignments for various anomaly-free 
discrete $\mathbb{Z}_{n}^R$ symmetries which are consistent with $SU(5)$ or 
$SO(10)$ unification (from Lee {\it et al.} Ref.~\cite{Lee:2011dya}).
}
\label{tab:R}
\end{center}
\end{table}
The virtues of the above discrete $R$-symmetries are multiple\cite{Hall:2002up,Chen:2012tia}.
\bi
\item They forbid the RPV superpotential terms, thus saving the proton from
  decay via dim-4 operators.
\item They forbid as well the dim-5 p-decay operators
  (which may be regenerated at the level of $\kappa\sim m_{3/2}/m_P$).
\item They allow for the usual Yukawa couplings needed for mass terms for SM fermions.
\item They allow for right-hand neutrino terms needed for a successful neutrino see-saw.
\item And they forbid the $\mu$ term.
  It is shown in Ref. \cite{Lee:2011dya} that this is not so for discrete gauge symmetries.
\item For high enough order, then the $\mathbb{Z}_n^R$ symmetries also
  offer a solution to the axion quality problem by forbidding non-renormalizable operators to sufficiently high order\cite{Lee:2011dya,Baer:2018avn,Bhattiprolu:2021rrj}. This can also be accomplished by sufficiently high-order discrete gauge symmetries; explicit $\mathbb{Z}_N$ examples with $N=11,12$ were given in Ref.~\cite{Babu:2002ic}, while a $\mathbb{Z}_{22}$ realization appears in the MBGW construction reviewed in Ref.~\cite{Bae:2019dgg}.
\ei

The $\mu$ term can be regenerated in SUSY via several two-extra-field models
dubbed {\it Base models I, II, III and IV} by \cite{Bhattiprolu:2021rrj}.
These are listed in Table \ref{tab:base}.
The $R$-charges of the $X$ and $Y$ fields depend on which $\mathbb{Z}_n^R$
is selected and which base model as well.
Full tabulations of the many choices are given in Ref. \cite{Baer:2025srs}.
\begin{table}[!htb]
\renewcommand{\arraystretch}{1.2}
\begin{center}
\begin{tabular}{c|cc}
base model & superpotential & PQ(X,Y) \\
\hline
$B_I$ (hyMSY) & $XYH_uH_d+X^3Y$ & $(-1,3)$ \\
$B_{II}$ (hyCCK/GSPQ) & $X^2H_uH_d+X^3Y$ & $(1,-3)$ \\
$B_{III}$ (hySPM) & $Y^2H_uH_d+X^3Y$ & $(-1/3,1)$ \\
$B_{IV}$ (MBGW) & $X^2H_uH_d+X^2Y^2$ & $(1,-1)$ \\
\hline
\end{tabular}
\caption{Four base models\cite{Bhattiprolu:2021rrj} along with
  associated PQ-sector superpotentials and PQ charges of $X$ and $Y$ fields.
}
\label{tab:base}
\end{center}
\end{table}

The scalar potential for each of the base models can be constructed
by augmenting the SUSY-conserving Lagrangian $F$-terms with the associated
soft SUSY breaking terms. In Ref. \cite{Murayama:1992dj,Bae:2014yta}, base model I is shown to develop
a non-zero minimum in the $\phi_X$ and $\phi_Y$ fields by radiatively
driving the $m_X^2$ term to negative values. However, the scalar potential
also develops a minimum via large soft terms of the form
$V\supset \left(\frac{f A_f\phi_X^2\phi_Y^2}{m_P}+h.c.\right)$.
Thus, due to SUSY breaking, the $\phi_{X,Y}$ fields develop vevs of order
$v_{X,Y}\sim \sqrt{m_{weak} m_P}$, thus also breaking the discrete $\mathbb{Z}_n^R$
symmetry. This then generates a $\mu$ term
\be
\mu \sim \lambda_\mu v_{X,Y}^2/m_P\sim m_{weak}
\ee
of the required magnitude!

\section{Origin of $U(1)_{PQ}$ as an accidental, approximate phase symmetry
in the $\mathbb{Z}_n^R$ augmented MSSM}
\label{sec:PQ}

At this point, we have constructed the MSSM as usual except that we
have replaced the imposition of RPC by the imposition of one of the
anomaly-free $\mathbb{Z}_n^R$ symmetries.
This improves upon the usual MSSM construction in that it
\begin{enumerate}
\item forbids the $\mu$ term,
\item forbids the various RPV terms and
\item forbids the dangerous dim-5 p-decay operators all the while
\item allowing for the usual superpotential Yukawa terms
and terms needed for imposing the neutrino see-saw.
\end{enumerate}
The $\mu$ term is regenerated as a consequence of SUSY breaking.

Now {\it a wonderful thing happens to the model before SUSY breaking occurs}.
In non-SUSY axion models, the origin of the global $U(1)_{PQ}$ is a
mystery awaiting a compelling solution\cite{DiLuzio:2020wdo} (as remarked in
Sec. 1).
In the case of the MSSM augmented by one of the $\mathbb{Z}_n^R$ symmetries--
and supplemented by any of the four base models--
the resultant model enjoys a global $U(1)_{PQ}$ phase invariance where
superfields transform as $\Phi \to e^{i\alpha q}\Phi$ (where $\Phi$ is a
generic left-chiral superfield), $q$ is the charge and $\alpha$ is
the global phase.
The superfield PQ charge assignments for each of the base models
are listed in Table \ref{tab:PQ} (normalized so that $q(H_{u,d})=-1$).
(For invariance under the global $U(1)_{PQ}$, it is straightforward to
check that for each of the allowed MSSM superpotential terms, the PQ
charges sum to zero.)

It is important to note here
that the global $U(1)$ emerges as an {\it accidental} global
phase invariance, and occurs as a consequence of the imposition of a
discrete $R$-symmetry along with the structure of each base model,
in much the same way as global $B$ and $L$ conservation emerge
accidentally in the SM as a consequence of gauge invariance.
This is most propitious since global symmetries are well-known to be not
fundamental, and are not respected by quantum gravity effects
(black-hole no-hair arguments and, more broadly, the expectation that exact global symmetries are absent in quantum gravity)\cite{Krauss:1988zc,Banks:2010zn}.
\begin{table}[!htb]
\renewcommand{\arraystretch}{1.2}
\begin{center}
\begin{tabular}{c|cccc}
multiplet & $B_I$ (hyMSY) & $B_{II}$ (hyCCK/GSPQ) & $B_{III}$ (hySPM) & $B_{IV}$ (MBGW) \\
\hline
$H_u$ & $-1$  & $-1$ & -1  & -1 \\
$H_d$ & $-1$  & $-1$ & -1  & -1 \\
$Q$   & $1/2$ & 1 & $1/2$  & 1 \\
$L$   & $1/2$ & 1 & $5/6$  & 1 \\
$U^c$ & $1/2$ & 0 & $1/2$  & 0 \\
$D^c$ & $1/2$ & 0 & $1/2$  & 0 \\
$E^c$ & $1/2$ & 0 & $1/6$  & 0 \\
$N^c$ & $1/2$ & 0 & $1/6$  & 0 \\
$X$   & $-1$  & 1 & $-1/3$ & 1 \\
$Y$   & $3$   & -3 & $1$   & -1 \\
\hline
\end{tabular}
\caption{Possible PQ charge assignments for various superfields of the
  four base models $B_I-B_{IV}$. We normalize the PQ charges so that
  $q(H_{u,d})=-1$.
}
\label{tab:PQ}
\end{center}
\end{table}

Upon imposing SUSY breaking, the $X$ and $Y$ fields
develop intermediate-scale vevs which break the $\mathbb{Z}_n^R$ symmetry
and generate a weak scale $\mu$ term (see {\it e.g.}
Ref's \cite{Baer:2018avn} and \cite{Baer:2025srs} for
explicit calculations).
But now the global $U(1)_{PQ}$ is also broken since $X$ and $Y$
carry PQ charges.
By Goldstone's theorem, a massless Goldstone boson (a combination of the
$X$ and $Y$ field phases) must develop which is identified as the QCD axion.
In our base models, it is realized that these are just supersymmetrized
versions of the DFSZ axion model\cite{Dine:1981rt,Zhitnitsky:1980tq},
and so the axion is of SUSY DFSZ type.
The axion develops a tiny mass due to non-perturbative effects and so
becomes a pseudo-Goldstone boson\cite{GrillidiCortona:2015jxo}.

\section{Some phenomenological consequences}
\label{sec:pheno}

We have focused here on the SUSY DFSZ axion emerging from the MSSM
construction which incorporates a KN solution to the $\mu$ problem.
The DFSZ axion has a domain wall number $N_{DW}=6$, so there is a danger that
domain walls may form in the early universe\cite{Sikivie:1982qv} leading to,
among other things, overproduction of dark matter.
We assume throughout that the PQ symmetry is broken during
inflation and not restored afterwards ($f_a\agt \max (T_R, H_I/2\pi )$),
so any domain wall structures are inflated away.

In the model considered here, a SUSY DFSZ axion always appears and so would
be one of the constituents of dark matter. We remind that the imposition of
a $\mathbb{Z}_n^R$ symmetry forbids the several $R$-parity violating
terms from the superpotential, so one might expect RPC and the LSP to provide
an additional DM candidate.
In Ref. \cite{Baer:2025oid,Baer:2025srs}, it is noted that higher dimensional operators of the form
\be
W_{NR}\supset X^pY^q \Phi\Phi\Phi /m_P^{p+q}
\ee
(where $\Phi$ denotes a generic visible-sector matter superfield, so that
$\Phi\Phi\Phi$ stands for any of the trilinear RPV combinations $LLE^c$, $LQD^c$ or
$U^cD^cD^c$) may be allowed, and have been tabulated in \cite{Baer:2025srs}.
For some cases, exact $R$-parity remains conserved, while for others
where $p+q\ge 3$, then $R$-parity is violated but with induced
RPV coefficients of order $(f_a/m_P)^{p+q}$ leading to an LSP with lifetime
longer than the age of the universe. Still other cases emerge
where the RPV operators have $p+q=1$ and are suppressed by
$(f_a/m_P)^1$ in which
case the LSPs (likely the lightest higgsino-like neutralino in natural SUSY models) would be produced as usual in the early universe, but would decay away
before or during the onset of Big Bang nucleosynthesis (BBN), leaving a SUSY universe
with all axion dark matter. This case may be in accord with recent null WIMP
search results from the LZ experiment\cite{LZ:2024zvo,LZ:2025igz}. In this case, the LSP
decay length was computed as typically at the km level\cite{Baer:2025srs}, so SUSY events
at colliders like LHC would still contain missing $E_T$ events, and the LSP
would decay well outside of the detector.
As noted in Ref. \cite{Baer:2025srs}, products of trilinear RPV couplings
must still respect the stringent proton decay bound
$\lambda_{11k}^\prime\lambda_{11k}^{\prime\prime}\alt 10^{-25}$, which may
require additional flavor-dependent suppression such as lepton triality.

The reduced $a\gamma\gamma$ coupling of the SUSY DFSZ axion makes haloscope searches
more challenging\cite{Bae:2017hlp};
nevertheless, modern haloscopes have now reached DFSZ sensitivity in selected mass
ranges\cite{ADMX:2021nhd,ADMX:2024xbv}, and the experimental program is advancing
rapidly\cite{Baryakhtar:2025jwh}.
In contrast to non-SUSY DFSZ models, now one must include charged higgsino
states in the $a\gamma\gamma$ triangle coupling, which leads to $E/N=6/3$
and an almost complete cancellation of coupling contributions.

\section{Conclusions}
\label{sec:conclude}

In this paper, we have pointed out that the axion solution to
the strong CP problem emerges accidentally and serendipitously
as a byproduct of solving the SUSY $\mu$ problem in the MSSM.
$N=1$ spacetime SUSY can emerge from string compactifications on Calabi-Yau
manifolds, and its breaking at the weak scale is supported by solving
the gauge hierarchy problem.
Discrete $R$ symmetries can arise as discrete
remnants from compactification of 10-d Lorentzian spacetime to 4-dimensions,
and their imposition provides the first step in solving the SUSY $\mu$ problem:
namely, forbidding the appearance of $\mu$ at scales far beyond the weak scale.
The coupling of the Higgs bilinear $H_uH_d$ to $R$-charged fields $X$ and $Y$
via non-renormalizable operators is in accord with the $\mathbb{Z}_n^R$
symmetries, but in this guise the theory develops an accidental global
$U(1)$ symmetry which can be identified as $U(1)_{PQ}$.
Then under SUSY breaking, the $X$ and $Y$ fields develop $\mathbb{Z}_n^R$
and PQ breaking vevs at an intermediate scale
(the cosmologically-favored sweet spot
$f_a\sim \sqrt{m_{weak}m_P}\sim 10^{10}-10^{12}$ GeV) in the KN mechanism.
The spontaneously broken global PQ then develops an axion as the associated
Goldstone boson. The $U(1)_{PQ}$ is also broken via anomalies, so the axion
develops its mass at scale $m_a\sim \Lambda_{QCD}^2/f_a$, and its coupling
to gluons allows the CP-violating Lagrangian terms that contribute to
the neutron EDM to relax towards zero, thus solving the strong CP problem.
In this sense, the axion solution
to the strong CP problem is a derived consequence of supersymmetrizing
the SM, and stabilizing its $\mu$ term via one of the
anomaly-free $\mathbb{Z}_n^R$ symmetries, and then regenerating it via
a Kim-Nilles operator.
This result may well be known to a number of researchers\cite{Kim:1983dt,Martin:2000eq,Lee:2011dya}, but it seems to be
hidden in the literature and certainly not appreciated by the wider
community.

{\it Acknowledgements:} 

V.B. gratefully acknowledges support from the U.S. Department of Energy,
Office of Science, Office of High Energy Physics,
under Award Number DE-SC0017647 and from the William F. Vilas Estate.
HB gratefully acknowledges support from the Avenir Foundation. D.S. gratefully acknowledges support from the European Research Council (ERC) under the European Union's Horizon 2020 research and innovation programme (Grant agreement No. 949451).

\bibliography{mssm}

@article{Nilles:2013lda,
    author = "Nilles, Hans Peter and Ramos-S{\'a}nchez, Sa{\'u}l and Ratz, Michael and Vaudrevange, Patrick K. S.",
    title = "{A note on discrete $R$ symmetries in $\mathbb{Z}_{6}$-II orbifolds with Wilson lines}",
    eprint = "1308.3435",
    archivePrefix = "arXiv",
    primaryClass = "hep-th",
    reportNumber = "DESY-13-143, TUM-HEP-901-13, FLAVOUR-EU-52-13",
    doi = "10.1016/j.physletb.2013.09.041",
    journal = "Phys. Lett. B",
    volume = "726",
    pages = "876--881",
    year = "2013"
}

@article{Nilles:2017heg,
    author = "Nilles, Hans Peter",
    title = "{Stringy Origin of Discrete R-symmetries}",
    eprint = "1705.01798",
    archivePrefix = "arXiv",
    primaryClass = "hep-ph",
    doi = "10.22323/1.292.0017",
    journal = "PoS",
    volume = "CORFU2016",
    pages = "017",
    year = "2017"
}

@article{Hinchliffe:1992ad,
    author = "Hinchliffe, Ian and Kaeding, Thomas",
    title = "{B+L violating couplings in the minimal supersymmetric Standard Model}",
    reportNumber = "LBL-32719",
    doi = "10.1103/PhysRevD.47.279",
    journal = "Phys. Rev. D",
    volume = "47",
    pages = "279--284",
    year = "1993"
}

@article{Hall:2002up,
    author = "Hall, Lawrence J. and Nomura, Yasunori and Pierce, Aaron",
    title = "{R symmetry and the mu problem}",
    eprint = "hep-ph/0204062",
    archivePrefix = "arXiv",
    reportNumber = "UCB-PTH-02-11, LBNL-49936",
    doi = "10.1016/S0370-2693(02)02043-9",
    journal = "Phys. Lett. B",
    volume = "538",
    pages = "359--365",
    year = "2002"
}

@article{DiLuzio:2020wdo,
    author = "Di Luzio, Luca and Giannotti, Maurizio and Nardi, Enrico and Visinelli, Luca",
    title = "{The landscape of QCD axion models}",
    eprint = "2003.01100",
    archivePrefix = "arXiv",
    primaryClass = "hep-ph",
    reportNumber = "DESY 20-036, DESY-20-036",
    doi = "10.1016/j.physrep.2020.06.002",
    journal = "Phys. Rept.",
    volume = "870",
    pages = "1--117",
    year = "2020"
}

@article{Banks:2010zn,
    author = "Banks, Tom and Seiberg, Nathan",
    title = "{Symmetries and Strings in Field Theory and Gravity}",
    eprint = "1011.5120",
    archivePrefix = "arXiv",
    primaryClass = "hep-th",
    doi = "10.1103/PhysRevD.83.084019",
    journal = "Phys. Rev. D",
    volume = "83",
    pages = "084019",
    year = "2011"
}

@article{LZ:2025igz,
    author = "Akerib, D. S. and others",
    collaboration = "LZ",
    title = "{Searches for Light Dark Matter and Evidence of Coherent Elastic Neutrino-Nucleus Scattering of Solar Neutrinos with the LUX-ZEPLIN (LZ) Experiment}",
    eprint = "2512.08065",
    archivePrefix = "arXiv",
    primaryClass = "hep-ex",
    month = "12",
    year = "2025"
}

@article{ADMX:2024xbv,
    author = "Goodman, C. and others",
    collaboration = "ADMX",
    title = "{ADMX Axion Dark Matter Bounds around 3.3{\,}{\,}{\ensuremath{\mu}}eV with Dine-Fischler-Srednicki-Zhitnitsky Discovery Ability}",
    eprint = "2408.15227",
    archivePrefix = "arXiv",
    primaryClass = "hep-ex",
    reportNumber = "FERMILAB-PUB-24-0602-V",
    doi = "10.1103/PhysRevLett.134.111002",
    journal = "Phys. Rev. Lett.",
    volume = "134",
    number = "11",
    pages = "111002",
    year = "2025"
}

@article{Baryakhtar:2025jwh,
    author = "Baryakhtar, Masha and Rosenberg, Leslie and Rybka, Gray",
    title = "{Searching for the QCD Dark Matter Axion}",
    eprint = "2504.10607",
    archivePrefix = "arXiv",
    primaryClass = "hep-ex",
    month = "4",
    year = "2025"
}

@inproceedings{Polchinski:2006gy,
    author = "Polchinski, Joseph",
    title = "{The Cosmological Constant and the String Landscape}",
    booktitle = "{23rd Solvay Conference in Physics: The Quantum Structure of Space and Time}",
    eprint = "hep-th/0603249",
    archivePrefix = "arXiv",
    pages = "216--236",
    month = "3",
    year = "2006"
}

@article{Baer:2020kwz,
    author = "Baer, Howard and Barger, Vernon and Salam, Shadman and Sengupta, Dibyashree and Sinha, Kuver",
    title = "{Status of weak scale supersymmetry after LHC Run 2 and ton-scale noble liquid WIMP searches}",
    eprint = "2002.03013",
    archivePrefix = "arXiv",
    primaryClass = "hep-ph",
    reportNumber = "OU-HEP-191231",
    doi = "10.1140/epjst/e2020-000020-x",
    journal = "Eur. Phys. J. ST",
    volume = "229",
    number = "21",
    pages = "3085--3141",
    year = "2020"
}

@article{Kim:2008hd,
    author = "Kim, Jihn E. and Carosi, Gianpaolo",
    title = "{Axions and the Strong CP Problem}",
    eprint = "0807.3125",
    archivePrefix = "arXiv",
    primaryClass = "hep-ph",
    doi = "10.1103/RevModPhys.82.557",
    journal = "Rev. Mod. Phys.",
    volume = "82",
    pages = "557--602",
    year = "2010",
    note = "[Erratum: Rev.Mod.Phys. 91, 049902 (2019)]"
}

@article{Jungman:1995df,
    author = "Jungman, Gerard and Kamionkowski, Marc and Griest, Kim",
    title = "{Supersymmetric dark matter}",
    eprint = "hep-ph/9506380",
    archivePrefix = "arXiv",
    reportNumber = "SU-4240-605, UCSD-PTH-95-02, IASSNS-HEP-95-14, CU-TP-677",
    doi = "10.1016/0370-1573(95)00058-5",
    journal = "Phys. Rept.",
    volume = "267",
    pages = "195--373",
    year = "1996"
}

@article{Martin:2000eq,
    author = "Martin, Stephen P.",
    title = "{Collider signals from slow decays in supersymmetric models with an intermediate scale solution to the mu problem}",
    eprint = "hep-ph/0005116",
    archivePrefix = "arXiv",
    reportNumber = "FERMILAB-PUB-00-097-T",
    doi = "10.1103/PhysRevD.62.095008",
    journal = "Phys. Rev. D",
    volume = "62",
    pages = "095008",
    year = "2000"
}

@article{Sikivie:2006ni,
    author = "Sikivie, Pierre",
    editor = "Kuster, Markus and Raffelt, Georg and Beltran, Berta",
    title = "{Axion Cosmology}",
    eprint = "astro-ph/0610440",
    archivePrefix = "arXiv",
    reportNumber = "UFIFT-HEP-06-16",
    doi = "10.1007/978-3-540-73518-2_2",
    journal = "Lect. Notes Phys.",
    volume = "741",
    pages = "19--50",
    year = "2008"
}

@article{Dreiner:1997uz,
    author = "Dreiner, Herbert K.",
    editor = "Kane, Gordon L.",
    title = "{An Introduction to explicit R-parity violation}",
    eprint = "hep-ph/9707435",
    archivePrefix = "arXiv",
    doi = "10.1142/9789814307505_0017",
    journal = "Adv. Ser. Direct. High Energy Phys.",
    volume = "21",
    pages = "565--583",
    year = "2010"
}

@article{Martin:1997ns,
    author = "Martin, Stephen P.",
    editor = "Kane, Gordon L.",
    title = "{A Supersymmetry primer}",
    eprint = "hep-ph/9709356",
    archivePrefix = "arXiv",
    reportNumber = "FERMILAB-PUB-97-425-T",
    doi = "10.1142/9789812839657_0001",
    journal = "Adv. Ser. Direct. High Energy Phys.",
    volume = "18",
    pages = "1--98",
    year = "1998"
}

@article{Dreiner:2005rd,
    author = "Dreiner, Herbi K. and Luhn, Christoph and Thormeier, Marc",
    title = "{What is the discrete gauge symmetry of the MSSM?}",
    eprint = "hep-ph/0512163",
    archivePrefix = "arXiv",
    doi = "10.1103/PhysRevD.73.075007",
    journal = "Phys. Rev. D",
    volume = "73",
    pages = "075007",
    year = "2006"
}

@article{LZ:2024zvo,
    author = "Aalbers, J. and others",
    collaboration = "LZ",
    title = "{Dark Matter Search Results from 4.2{\,}{\,}Tonne-Years of Exposure of the LUX-ZEPLIN (LZ) Experiment}",
    eprint = "2410.17036",
    archivePrefix = "arXiv",
    primaryClass = "hep-ex",
    reportNumber = "FERMILAB-PUB-24-0796-V",
    doi = "10.1103/4dyc-z8zf",
    journal = "Phys. Rev. Lett.",
    volume = "135",
    number = "1",
    pages = "011802",
    year = "2025"
}

@article{Abel:2020pzs,
    author = "Abel, C. and others",
    title = "{Measurement of the Permanent Electric Dipole Moment of the Neutron}",
    eprint = "2001.11966",
    archivePrefix = "arXiv",
    primaryClass = "hep-ex",
    doi = "10.1103/PhysRevLett.124.081803",
    journal = "Phys. Rev. Lett.",
    volume = "124",
    number = "8",
    pages = "081803",
    year = "2020"
}

@article{Peccei:1977hh,
    author = "Peccei, R. D. and Quinn, Helen R.",
    title = "{CP Conservation in the Presence of Instantons}",
    reportNumber = "ITP-568-STANFORD",
    doi = "10.1103/PhysRevLett.38.1440",
    journal = "Phys. Rev. Lett.",
    volume = "38",
    pages = "1440--1443",
    year = "1977"
}

@article{Peccei:1977ur,
    author = "Peccei, R. D. and Quinn, Helen R.",
    title = "{Constraints Imposed by CP Conservation in the Presence of Instantons}",
    reportNumber = "ITP-572-STANFORD",
    doi = "10.1103/PhysRevD.16.1791",
    journal = "Phys. Rev. D",
    volume = "16",
    pages = "1791--1797",
    year = "1977"
}

@article{Weinberg:1977ma,
    author = "Weinberg, Steven",
    title = "{A New Light Boson?}",
    reportNumber = "HUTP-77/A074",
    doi = "10.1103/PhysRevLett.40.223",
    journal = "Phys. Rev. Lett.",
    volume = "40",
    pages = "223--226",
    year = "1978"
}

@article{Wilczek:1977pj,
    author = "Wilczek, Frank",
    title = "{Problem of Strong  $P$  and  $T$  Invariance in the Presence of Instantons}",
    reportNumber = "Print-77-0939 (COLUMBIA)",
    doi = "10.1103/PhysRevLett.40.279",
    journal = "Phys. Rev. Lett.",
    volume = "40",
    pages = "279--282",
    year = "1978"
}

@article{Sikivie:1982qv,
    author = "Sikivie, P.",
    title = "{Of Axions, Domain Walls and the Early Universe}",
    reportNumber = "UFTP-82-3",
    doi = "10.1103/PhysRevLett.48.1156",
    journal = "Phys. Rev. Lett.",
    volume = "48",
    pages = "1156--1159",
    year = "1982"
}

@article{Dine:1981rt,
    author = "Dine, Michael and Fischler, Willy and Srednicki, Mark",
    title = "{A Simple Solution to the Strong CP Problem with a Harmless Axion}",
    reportNumber = "Print-81-0320 (IAS,PRINCETON)",
    doi = "10.1016/0370-2693(81)90590-6",
    journal = "Phys. Lett. B",
    volume = "104",
    pages = "199--202",
    year = "1981"
}

@article{Zhitnitsky:1980tq,
    author = "Zhitnitsky, A. R.",
    title = "{On Possible Suppression of the Axion Hadron Interactions. (In Russian)}",
    journal = "Sov. J. Nucl. Phys.",
    volume = "31",
    pages = "260",
    year = "1980"
}

@article{Ibanez:1991pr,
    author = "Ibanez, Luis E. and Ross, Graham G.",
    title = "{Discrete gauge symmetries and the origin of baryon and lepton number conservation in supersymmetric versions of the standard model}",
    reportNumber = "CERN-TH-6111-91",
    doi = "10.1016/0550-3213(92)90195-H",
    journal = "Nucl. Phys. B",
    volume = "368",
    pages = "3--37",
    year = "1992"
}

@article{Bae:2019dgg,
    author = "Bae, Kyu Jung and Baer, Howard and Barger, Vernon and Sengupta, Dibyashree",
    title = "{Revisiting the SUSY $\mu$ problem and its solutions in the LHC era}",
    eprint = "1902.10748",
    archivePrefix = "arXiv",
    primaryClass = "hep-ph",
    reportNumber = "CTPU-PTC-19-06",
    doi = "10.1103/PhysRevD.99.115027",
    journal = "Phys. Rev. D",
    volume = "99",
    number = "11",
    pages = "115027",
    year = "2019"
}

@article{Bae:2014rfa,
    author = "Bae, Kyu Jung and Baer, Howard and Lessa, Andre and Serce, Hasan",
    title = "{Coupled Boltzmann computation of mixed axion neutralino dark matter in the SUSY DFSZ axion model}",
    eprint = "1406.4138",
    archivePrefix = "arXiv",
    primaryClass = "hep-ph",
    reportNumber = "OU-HEP-140531",
    doi = "10.1088/1475-7516/2014/10/082",
    journal = "JCAP",
    volume = "10",
    pages = "082",
    year = "2014"
}

@article{Krauss:1988zc,
    author = "Krauss, Lawrence M. and Wilczek, Frank",
    title = "{Discrete Gauge Symmetry in Continuum Theories}",
    reportNumber = "YCTP-P26-88, NSF-ITP-88-187",
    doi = "10.1103/PhysRevLett.62.1221",
    journal = "Phys. Rev. Lett.",
    volume = "62",
    pages = "1221",
    year = "1989"
}

@article{Nilles:2009yd,
    author = "Nilles, Hans Peter and Ramos-Sanchez, Saul and Vaudrevange, Patrick K. S.",
    editor = "Alverson, George and Nelson, Brent and Nath, Pran",
    title = "{Local Grand Unification and String Theory}",
    eprint = "0909.3948",
    archivePrefix = "arXiv",
    primaryClass = "hep-th",
    reportNumber = "DESY-09-147, LMU-ASC-40-09",
    doi = "10.1063/1.3327561",
    journal = "AIP Conf. Proc.",
    volume = "1200",
    number = "1",
    pages = "226--234",
    year = "2010"
}

@inproceedings{Buchmuller:2005sh,
    author = "Buchmuller, Wilfried and Hamaguchi, Koichi and Lebedev, Oleg and Ratz, Michael",
    title = "{Local grand unification}",
    booktitle = "{GUSTAVOFEST: Symposium in Honor of Gustavo C. Branco: CP Violation and the Flavor Puzzle}",
    eprint = "hep-ph/0512326",
    archivePrefix = "arXiv",
    reportNumber = "DESY-05-260",
    pages = "143--156",
    month = "12",
    year = "2005"
    }

@article{Ratz:2007my,
    author = "Ratz, Michael",
    title = "{Notes on Local Grand Unification}",
    eprint = "0711.1582",
    archivePrefix = "arXiv",
    primaryClass = "hep-ph",
    reportNumber = "TUM-HEP-07-678",
    doi = "10.24532/soken.116.1_A56",
    journal = "Soryushiron Kenkyu Electron.",
    volume = "116",
    pages = "A56--A76",
    year = "2008"
}

@article{Baer:2018avn,
    author = "Baer, Howard and Barger, Vernon and Sengupta, Dibyashree",
    title = "{Gravity safe, electroweak natural axionic solution to strong $CP$ and SUSY $\mu$ problems}",
    eprint = "1810.03713",
    archivePrefix = "arXiv",
    primaryClass = "hep-ph",
    reportNumber = "OU-HEP-180930",
    doi = "10.1016/j.physletb.2019.01.007",
    journal = "Phys. Lett. B",
    volume = "790",
    pages = "58--63",
    year = "2019"
}

@article{Bhattiprolu:2021rrj,
    author = "Bhattiprolu, Prudhvi N. and Martin, Stephen P.",
    title = "{High-quality axions in solutions to the {\ensuremath{\mu}} problem}",
    eprint = "2106.14964",
    archivePrefix = "arXiv",
    primaryClass = "hep-ph",
    doi = "10.1103/PhysRevD.104.055014",
    journal = "Phys. Rev. D",
    volume = "104",
    number = "5",
    pages = "055014",
    year = "2021"
}

@article{Babu:2002ic,
    author = "Babu, K. S. and Gogoladze, Ilia and Wang, Kai",
    title = "{Stabilizing the axion by discrete gauge symmetries}",
    eprint = "hep-ph/0212339",
    archivePrefix = "arXiv",
    reportNumber = "OSU-HEP-02-18",
    doi = "10.1016/S0370-2693(03)00411-8",
    journal = "Phys. Lett. B",
    volume = "560",
    pages = "214--222",
    year = "2003"
}

@article{Lee:2011dya,
    author = "Lee, Hyun Min and Raby, Stuart and Ratz, Michael and Ross, Graham G. and Schieren, Roland and Schmidt-Hoberg, Kai and Vaudrevange, Patrick K. S.",
    title = "{Discrete R symmetries for the MSSM and its singlet extensions}",
    eprint = "1102.3595",
    archivePrefix = "arXiv",
    primaryClass = "hep-ph",
    reportNumber = "TUM-HEP-793-11, LMU-ASC-06-11, OHSTPY-HEP-T-11-001, CERN-PH-TH-2011-022, OUTP-11-33P",
    doi = "10.1016/j.nuclphysb.2011.04.009",
    journal = "Nucl. Phys. B",
    volume = "850",
    pages = "1--30",
    year = "2011"
}

@article{Baer:2024fgd,
    author = "Baer, Howard and Barger, Vernon and Bolich, Jessica and Zhang, Kairui",
    title = "{Implications of Higgs mass for hidden sector SUSY breaking}",
    eprint = "2412.15356",
    archivePrefix = "arXiv",
    primaryClass = "hep-ph",
    reportNumber = "OU-HEP-241225",
    doi = "10.1103/PhysRevD.111.095019",
    journal = "Phys. Rev. D",
    volume = "111",
    number = "9",
    pages = "095019",
    year = "2025"
}

@article{Green:1984sg,
    author = "Green, Michael B. and Schwarz, John H.",
    title = "{Anomaly Cancellation in Supersymmetric D=10 Gauge Theory and Superstring Theory}",
    reportNumber = "CALT-68-1182",
    doi = "10.1016/0370-2693(84)91565-X",
    journal = "Phys. Lett. B",
    volume = "149",
    pages = "117--122",
    year = "1984"
}

@article{GrillidiCortona:2015jxo,
    author = "Grilli di Cortona, Giovanni and Hardy, Edward and Pardo Vega, Javier and Villadoro, Giovanni",
    title = "{The QCD axion, precisely}",
    eprint = "1511.02867",
    archivePrefix = "arXiv",
    primaryClass = "hep-ph",
    doi = "10.1007/JHEP01(2016)034",
    journal = "JHEP",
    volume = "01",
    pages = "034",
    year = "2016"
}

@article{Baer:2025srs,
    author = "Baer, Howard and Barger, Vernon and Bolich, Jessica and Sengupta, Dibyashree and Zhang, Kairui",
    title = "{Aspects of the WIMP quality problem and R-parity violation in natural supersymmetry with all axion dark~matter}",
    eprint = "2505.09785",
    archivePrefix = "arXiv",
    primaryClass = "hep-ph",
    reportNumber = "OU-HEP-250509",
    doi = "10.1088/1475-7516/2025/10/072",
    journal = "JCAP",
    volume = "10",
    pages = "072",
    year = "2025"
}

@article{Baer:2025oid,
    author = "Baer, Howard and Barger, Vernon and Sengupta, Dibyashree and Zhang, Kairui",
    title = "{All axion dark matter from supersymmetric models}",
    eprint = "2502.06955",
    archivePrefix = "arXiv",
    primaryClass = "hep-ph",
    reportNumber = "OU-HEP-250204",
    doi = "10.1103/1n3m-g69m",
    journal = "Phys. Rev. D",
    volume = "111",
    number = "11",
    pages = "L111702",
    year = "2025"
}

@article{Chen:2012tia,
    author = "Chen, Mu-Chun and Fallbacher, Maximilian and Ratz, Michael",
    editor = "Szczerbinska, Barbara and Babu, Kaladi and Balantekin, Baha and Dutta, Bhaskar and Mohapatra, Rabindra N.",
    title = "{Supersymmetric unification and R symmetries}",
    eprint = "1211.6247",
    archivePrefix = "arXiv",
    primaryClass = "hep-ph",
    reportNumber = "UCI-TR-2012-19, TUM-HEP-869-12, FLAVOR-EU-32, CETUP-016",
    doi = "10.1142/S0217732312300443",
    journal = "Mod. Phys. Lett. A",
    volume = "27",
    pages = "1230044",
    year = "2012"
}

@article{Kim:1983dt,
    author = "Kim, Jihn E. and Nilles, Hans Peter",
    title = "{The mu Problem and the Strong CP Problem}",
    reportNumber = "UGVA-DPT 1983/10-410",
    doi = "10.1016/0370-2693(84)91890-2",
    journal = "Phys. Lett. B",
    volume = "138",
    pages = "150--154",
    year = "1984"
}

@article{Giudice:1988yz,
    author = "Giudice, G. F. and Masiero, A.",
    title = "{A Natural Solution to the mu Problem in Supergravity Theories}",
    reportNumber = "SISSA-17/88/EP",
    doi = "10.1016/0370-2693(88)91613-9",
    journal = "Phys. Lett. B",
    volume = "206",
    pages = "480--484",
    year = "1988"
}

@article{Ellwanger:2009dp,
    author = "Ellwanger, Ulrich and Hugonie, Cyril and Teixeira, Ana M.",
    title = "{The Next-to-Minimal Supersymmetric Standard Model}",
    eprint = "0910.1785",
    archivePrefix = "arXiv",
    primaryClass = "hep-ph",
    reportNumber = "LPT-ORSAY-09-76, CFTP-09-032, LPTA-09-066",
    doi = "10.1016/j.physrep.2010.07.001",
    journal = "Phys. Rept.",
    volume = "496",
    pages = "1--77",
    year = "2010"
}

@article{Murayama:1992dj,
    author = "Murayama, H. and Suzuki, H. and Yanagida, T.",
    title = "{Radiative breaking of Peccei-Quinn symmetry at the intermediate mass scale}",
    reportNumber = "TU-407",
    doi = "10.1016/0370-2693(92)91397-R",
    journal = "Phys. Lett. B",
    volume = "291",
    pages = "418--425",
    year = "1992"
}

@article{Bae:2014yta,
    author = "Bae, Kyu Jung and Baer, Howard and Serce, Hasan",
    title = "{Natural little hierarchy for SUSY from radiative breaking of the Peccei-Quinn symmetry}",
    eprint = "1410.7500",
    archivePrefix = "arXiv",
    primaryClass = "hep-ph",
    reportNumber = "FTPI-MINN-14-36",
    doi = "10.1103/PhysRevD.91.015003",
    journal = "Phys. Rev. D",
    volume = "91",
    number = "1",
    pages = "015003",
    year = "2015"
}

@book{Baer:2006rs,
    author = "Baer, H. and Tata, X.",
    title = "{Weak scale supersymmetry: From superfields to scattering events}",
    isbn = "978-0-521-29031-9, 978-0-511-19011-7, 978-0-521-29031-9, 978-0-521-85786-4",
    publisher = "Cambridge University Press",
    month = "5",
    year = "2006"
}

@article{Bae:2017hlp,
    author = "Bae, Kyu Jung and Baer, Howard and Serce, Hasan",
    title = "{Prospects for axion detection in natural SUSY with mixed axion-higgsino dark matter: back to invisible?}",
    eprint = "1705.01134",
    archivePrefix = "arXiv",
    primaryClass = "hep-ph",
    reportNumber = "CTPU-17-16",
    doi = "10.1088/1475-7516/2017/06/024",
    journal = "JCAP",
    volume = "06",
    pages = "024",
    year = "2017"
}

@article{ADMX:2021nhd,
    author = "Bartram, C. and others",
    collaboration = "ADMX",
    title = "{Search for Invisible Axion Dark Matter in the 3.3{\textendash}4.2{\,}{\,}{\ensuremath{\mu}}eV Mass Range}",
    eprint = "2110.06096",
    archivePrefix = "arXiv",
    primaryClass = "hep-ex",
    reportNumber = "FERMILAB-PUB-21-774-DI-PPD-SQMS",
    doi = "10.1103/PhysRevLett.127.261803",
    journal = "Phys. Rev. Lett.",
    volume = "127",
    number = "26",
    pages = "261803",
    year = "2021"
}
\bibliographystyle{elsarticle-num}

\end{document}